%% file: main.tex
\documentclass[10pt,aps,pra,superscriptaddress,nofootinbib]{revtex4-2}

\pdfoutput=1
\input{macros}

\begin{document}
	
\title{Tomography of clock signals using the simplest possible reference}

\author{Nuriya Nurgalieva}
\email{nuriya@squids.ch}
\affiliation{Institute for Theoretical Physics, ETH Z\"urich, 8093 Z\"urich, Switzerland}
\affiliation{Department of Physics, University of Zurich, 8057 Zurich, Switzerland}

\author{Ralph Silva}
\email{ralphfsilva@gmail.com}
\affiliation{Institute for Theoretical Physics, ETH Z\"urich, 8093 Z\"urich, Switzerland}

\author{Renato Renner}
\affiliation{Institute for Theoretical Physics, ETH Z\"urich, 8093 Z\"urich, Switzerland}

\begin{abstract}
\onecolumngrid
We show that finite physical clocks always have well-behaved signals, namely that every waiting-time distribution generated by a physical process on a system of finite size is guaranteed to be bounded by a decay envelope. Following this consideration, we show that one can reconstruct the distribution using only operationally available information, namely, that of the ordering of the ticks of one clock with the respect to those of another clock (which we call the reference), and that the simplest possible reference clock --- a Poisson process --- suffices.

\end{abstract}

\maketitle

\epigraph{Time and I have quarrelled. All hours are midnight now. I had a clock and a watch, but I destroyed them both. I could not bear the way they mocked me.}{\textit{Jonathan Strange \& Mr Norrell} \\ Suzanna Clark}

\twocolumngrid
\tableofcontents
\hrulefill
\bigskip

\section{Introduction}
\label{sec:introduction}
\input{1_introduction}

\section{Finite clocks always have well behaved signals}
\label{sec:finite}
\input{2_finite}

\section{A Poisson process: the simplest ticking clock}
\label{sec:poisson}
\input{3_poisson}

\section{Reconstruction of the characteristic function of an unknown clock}
\label{sec:construction}
\input{4_fish}

\bigskip

\section{Finite sampling: frequency as a resource}
\label{sec:frequency}
\input{5_frequency}

\section{Reconstructing a non i.i.d. clock}
\label{sec:nonreset}
\input{6_nonreset}

\section{Discussion \& Outlook}
\label{sec:conclusions}
\input{8_conclusions}

\section*{Acknowledgements}

The authors thank Henrik Wilming for interesting discussions. We acknowledge support from the Swiss National Science Foundation (SNSF) through project No. 200021\_188541 and through the National Centre of Competence in Research SwissMAP. R.S.\ further acknowledges funding from the Swiss National Science Foundation via an Ambizione grant PZ00P2\_185986.

\newpage
\onecolumngrid
\bibliographystyle{unsrtnat}

\input{bibliography.bbl}\onecolumngrid
\newpage
\appendix

\section*{Appendix}
\input{appendix}

\end{document}

%% file: macros.tex
\usepackage{amsmath}
\usepackage[english]{babel}
\usepackage[T1]{fontenc}
\usepackage[colorlinks,citecolor=blue,linkcolor=blue]{hyperref}
\usepackage[utf8]{inputenc}
\usepackage[dvipsnames]{xcolor}
\usepackage{comment}
\usepackage[backgroundcolor=green,textwidth=1.5cm,textsize=tiny]{todonotes}

\usepackage{smartdiagram}

\usepackage{amsfonts}
\usepackage{amsthm}
\usepackage{braket}
\usepackage{dsfont}
\usepackage[dvipsnames]{xcolor}
\usepackage{epigraph}

\theoremstyle{plain}

\newtheorem{exit-strategy}{Exit strategy}

\theoremstyle{definition}

\newtheorem*{axiom*}{Axiom}

\theoremstyle{remark}

\DeclareMathOperator{\Tr}{Tr}

\newcommand{\stirlingone}[2]{\genfrac{[}{]}{0pt}{}{#1}{#2}}
\newcommand{\stirlingtwo}[2]{\genfrac{\{}{\}}{0pt}{}{#1}{#2}}

\newcommand*{\tick}{\mathrm{tick}}

%% file: 1_introduction.tex
\paragraph{Motivation.}

Clocks come in various shapes and sizes, but the most recognisable are those that `tick', that is to say, they occasionally emit a clearly observable signal to indicate that a particular unit of time has passed, be it a second, an hour or any other time interval. By doing so, the time of an observer is split into segments, each marked by a different number of ticks counted by the clock.

An ideal clock would mark these intervals with perfect regularity, each tick arriving after exactly the same change in the background time. In reality, no clock is able to do so, whether for reasons of finite size~\cite{rankovic_quantum_2015,yang_ultimate_2020,woods_quantum_2022}, finite resources~\cite{erker_autonomous_2017}, or just experimental imperfections. For a non-ideal clock, the timing of the ticks is not a fixed stable constant, but rather a fluctuating quantity. Mathematically we quantify this by speaking of the \textit{probability density function} of the \textit{time of arrival} of the ticks, a function $\Pr(T_1=t_1,T_2=t_2,...)$ that measures the likelihood that the first tick $T_1$ happened at $t_1$, the second at $t_2$ and so on. For a clock whose ticks form an i.i.d.\ sequence, this is determined by the \textit{waiting-time distribution} of a single tick, $\Pr(T_n - T_{n-1}=t)$. More generally, as long as the ticks are identically distributed, the asymptotic behaviour of the clock is encapsulated in its \textit{full counting statistics}, namely the moments of the number of ticks of the clock for large enough time $t$.

When we observe a clock, however, we cannot observe these distributions directly as we have no access to $t$: the background time that our systems are dancing to the tune of. The only manner in which we can access $t$ is via a clock itself --- thus the only way for us to observe the ticks of one clock is to compare them to those of another. Thus arises a natural question: is it possible to use the \textit{operationally available information} --- that of the ordering of the ticks of one clock with respect to those of another reference clock --- to reconstruct the \textit{full abstract information} of the first clock, i.e.\ its probability density function?

\paragraph{Setting.}

In this paper, we show that for a clock whose ticks form an identically distributed sequence, we can indeed reconstruct the distribution by using arguably the simplest possible reference clock: an i.i.d.\ Poisson process. If the ticks of the clock of interest are also independent --- so that its probability density function for many ticks is determined by that of a single tick --- then we recover the waiting-time distribution of a single tick.\footnote{Also referred as the \textit{delay function} in some literature.} More generally, we can always recover the full counting statistics of the clock. 

The method of reconstruction works by determining the moments of the waiting-time distribution, and thus relies on them being well-behaved in a certain sense. In that light, we also prove here that every waiting-time distribution generated by a physical process on a system of finite size (as measured by its Hilbert space) is guaranteed to fulfill this requirement, and provide a sufficient criterion for the infinite-dimensional case by considering the Lindbladian dynamics of the system.


\paragraph{Contribution.}

Our result is a key conceptual step in the operational characterisation of clocks, a `tomography of time-signals' so to speak. It shows that
even the simplest clocks are good references given enough measurements; that describing a clock signal w.r.t.\ background time does make sense since we can operationally access it; and finally --- unsurprisingly --- that the frequency of the reference together with the memory of multiple measurements are resources in this process.

\paragraph{Structure.}

The paper is structured as follows. First, in Sec.~\ref{sec:finite} we clarify what we mean by by a clock and show that all clocks whose dynamics are generated by a finite state space have well-behaved delay functions. In Sec.~\ref{sec:poisson} we provide both mathematical and physical reasons for regarding the Poisson process as the simplest and most imprecise reference clock. This is followed in Sec.~\ref{sec:construction} by the method of reconstruction of the moments and characteristic function, and Sec.~\ref{sec:frequency} discusses the rate of convergence in the finite sampling case, based on the rate of the Poisson reference. Sec.~\ref{sec:nonreset} considers the reconstruction of non-iid clocks, before we conclude in Sec.~\ref{sec:conclusions}.

%% file: 2_finite.tex
\paragraph{Modelling the dynamics of a clock.} To understand the ticks generated by an imperfect clock, we have to model both the ticks and the physical system that generates them. This is customarily done via a bipartite structure: one has an internal clockwork ($C$) that is coupled to a classical integer-labelled register ($T$), the latter of which indicates how many ticks the clock has recorded. Any measurement of the observer is only upon $T$, the internal clockwork is left undisturbed. Crucially, while the register is strictly classical, the clockwork may be quantum-mechanical in nature. For a deeper discussion of this model and how it arises among time-keeping models in general, see~\cite{Silva2023}. We summarise the central points here.

At any given time, the information about the state of the clockwork and register is encoded in the following statistical ensemble or density matrix upon their joint Hilbert space $\mathcal{H}_C \otimes \mathcal{H}_T$:
\begin{align}\label{eq:bipartite}
    \tilde{\rho}_{CT} (t) &= \sum_n \rho^{(n)}_C (t) \otimes \ket{n}\!\bra{n}_T,
\end{align}
where the $\rho_C^{(n)}(t)$ are semi-positive matrices whose sum is normalised. The information solely related to ticks is encoded only upon the register, this is obtained by tracing out (ignoring) the state of the clockwork:
\begin{align}\label{eq:numberofticks}
    \eta_T(t) &= \sum_n \Tr_C \left[ \rho^{(n)}_C (t) \right] \ket{n}\!\bra{n}_T \equiv \left\{ \Pr(n|t) \right\}_n.
\end{align}
The distribution is generally not deterministic, reflecting the uncertainty in the number of recorded ticks of an imperfect clock.

Furthermore, it is assumed that the register evolves \emph{serially} and \emph{irreversibly}, i.e.\ that the register state can only move one step at a time from $n$ to $n+1$. Both of these simplifications suggest themselves directly from the behaviour of real clocks that we observe; we employ them in this work. Note that both can be relaxed for more general models, in particular the latter assumption of irreversibility when working with entropic clocks.

One can describe the behaviour of the ticks of the clock by their \emph{time of arrival}, i.e.\ the moment in time $T_n$ that the register state changes from $n-1$ to $n$. For imperfect clocks, these are not determined times; rather, one has a probability distribution for them, denoted by $\Pr(T_n = t)$. One can shift between the time-of-arrival picture and of the number-of-ticks picture above \eqref{eq:numberofticks} via the relation:
\begin{align}\label{eq:timeofarrival}
	\Pr(T_n = t) &= - \frac{d}{dt} \sum_{m=0}^{n-1} \Pr(m|t).
\end{align}

\paragraph{Self-contained clocks and the assumption of self-timing.} It is customary when describing the dynamics of a ticking clock to take it to be \emph{self-timed}, i.e., that its dynamics is time-independent and dependent only upon its own state. This ensures that all of the resources and time-keeping of the clock are contained within its description rather than arising from an external hidden clock. Under this assumption, the dynamics can be simply expressed as arising from a Lindbladian generator $\tilde{L}_{CT}$, albeit one that maintains the bipartite structure from \eqref{eq:bipartite}, in particular the classical nature of the tick register.

\paragraph{Reset clocks: i.i.d.\ ticks.} We take as a starting point the case of \emph{reset} clocks (equivalently \emph{i.i.d.}\ clocks). These have the property that the moment the clock ticks the clockwork returns to a particular fixed state regardless of the dynamics leading up to the tick. This ensures that the process of each tick is independent of the previous one as well as identical. As a result, the distribution of the time of arrival of every tick is entirely determined by that of a single tick, that of the clock when it begins in the fixed post-tick state (referred to as the reset state), via the $n$-fold convolution:
\begin{align}
	\Pr(T_n = t) &= \omega (t)^{\circ n},
\end{align}
where $\omega(t)$ is the \emph{waiting-time distribution} of a single tick (alternatively, as the \emph{delay function}).

\paragraph{Moments and the moment generating function.} One manner of recovering the waiting-time distribution $\omega(t)$ of an unknown clock is by finding its moments $M_j$:
\begin{align}
    M_j &= \int_0^\infty t^j \omega(t) dt,
\end{align}
where $j \in \mathbb{Z}^+$. If the distribution is well-behaved in a certain sense, then it is uniquely determined by the full set of moments $\{M_j\}$. In particular, the relation is one-to-one in the case that the \textit{moment generating function} $\mathcal{M}$ is well-defined in a finite region. More precisely, given
\begin{align}
    \mathcal{M} (x) &= \int_0^\infty e^{tx} \omega(t) dt,
\end{align}
the distribution $\omega$ is uniquely determined by its moments $M_j$ if there exists some $a>0$ such that $\mathcal{M}(x)$ is finite for $x<a$.

\paragraph{The moment generating function of finite clocks.} Our first result is that for clocks whose dynamics are generated by a finite state space --- so that $\mathcal{H}_C$ can be taken to be of finite size --- and which are guaranteed to tick eventually, the moment generating function is well-defined in a finite neighbourhood. We outline the argument here, the full proof is detailed in Appendix \ref{app:finite}. We prove the statement by showing that $\omega(t)$ falls under an exponential envelope $e^{-at}$ for some $a>0$, so that $\mathcal{M}(x)$ exists for $x<a$.

\textbf{Proof outline.} Under Lindbladian evolution the state of the clock and register at any time is the exponential $e^{\tilde{\mathcal{L}} t} [\tilde{\rho}_0]$, where $\tilde{\rho}_0$ is an arbitrary initial state. The generator is infinite-dimensional --- thanks to the register --- however, as we are only interested in a single tick of the clock, we can construct a reduced scenario wherein the dynamics after the first tick is ignored, giving us a finite state space of size $d+1$, where $d$ is the dimension of the clock Hilbert space and the extra space is to denote the clock having ticked, a state we label as $\ket{\tick}$. The reduced Lindbladian $\mathcal{L}^\prime$ is also finite and has the same dynamics for the first tick as the original.

The eigenvalues of the generator are either zero (corresponding to steady states) or negative (corresponding to decaying states). If the clock is guaranteed to tick eventually, then the only steady state is $\ket{\tick}$. One can now express the difference between the initial state of the clock and the steady state in terms of the eigenmatrices of $\mathcal{L}^\prime$:\footnote{If the generator is not diagonalisable then one has generalised eigenmatrices. The proof in Appendix \ref{app:finite} deals with this general case.}
\begin{align}
    \rho_0^\prime - \ket{\tick}\!\bra{\tick} &= \sum_i c_i \sigma_i.
\end{align}
We argue that every eigenmatrix appearing in the above sum must correspond to an eigenvalue with a strictly negative real part; this follows by applying the evolution operator and taking $t\to \infty$. The left side goes to zero as $\rho_0^\prime \to \ket{\tick}\!\bra{\tick}$ while $\ket{\tick}\!\bra{\tick}$ is preserved. The right side gains exponentials $e^{\lambda_i t}$ where $\lambda_i$ are the eigenvalues of $\mathcal{L}^\prime$. For this to go to zero, every $\lambda_i$ must have a negative real part.

By picking the eigenvalue with real part closest to zero, we obtain a lower bound on the decay rate of the above expression. This, in turn, allows us to obtain the same lower bound on the waiting-time distribution. From this, we can show that the moment generating function exists for $x<a$ where $a = - \max \Re(\lambda_i) > 0$ (the zero eigenvalue is excluded in the maximisation).

%% file: 3_poisson.tex
Next, we turn to the reference, which we pick to be arguably the simplest and most widely available resource: a Poisson process. Viewed as a clock, it is specified by the Poisson distribution of rate $\gamma$: in a time interval of $t$ the probability distribution of the number of ticks seen from this clock is
\begin{align}
    \Pr ( N_R = n | t ) &= \frac{(\gamma t)^n e^{-\gamma t}}{n!}.
\end{align}

An equivalent description of a Poisson process is via its waiting-time distribution: the probability density function of the time between one tick and the next is the exponential decay:
\begin{align}
    \Pr( T_n - T_{n-1} = t) &= \gamma e^{-\gamma t}.
\end{align}
The most distinctive example of a Poisson process is the decay of a radioactive atom.

The Poisson process has the property of being \textit{memoryless} in the sense that in an infinitesimal interval of time $dt$, the probability of a tick being seen is $\gamma dt$ and is entirely independent of the history of ticks and thus there are no correlations across time. This is in contrast to a perfectly regular clock whose probability density of ticking is highly correlated with the history of ticks.

The simplicity of the Poisson process is also seen with respect to the precision measure $P = \mu^2/\sigma^2$, where $\mu$ is the first moment of the waiting-time and $\sigma^2$ is the variance of the same. This measure derives from the question: how many ticks of an i.i.d.\ clock (which the Poisson process is) must we see before the variance of the ticks is as large as the expected interval between them? The better the clock the higher $P$ is expected to be. For the Poisson process, we have $P=1$ only.

We can also argue that the Poisson process is trivial from a complexity viewpoint: one can describe its dynamics without the need of a clockwork, i.e., its time evolution depends only upon the  entry in the register~$T$:
\begin{align}
    \eta_T &= \sum_{n=0}^\infty \Pr(n|t) \ket{n}\!\bra{n}_T, \\
    \frac{d}{dt} \eta = \mathcal{L} \left[ \eta \right] &= \gamma \left( \Gamma \eta \Gamma^\dagger - \frac{1}{2} \left\{ \Gamma^\dagger \Gamma, \eta \right\} \right), \\
    \text{where} \quad \Gamma &= \sum_0^\infty \ket{n+1}\!\bra{n}.
\end{align}

Thus, from the property of being memoryless, having minimal precision, and the emergence in systems of minimal complexity, the Poisson process suggests itself as the simplest possible clock. And yet, as we proceed to demonstrate, it makes a perfectly good reference.

The simplicity of the Poisson process allows for a useful property: increasing the effective rate of the process by taking many together. If one has ticks produced by two processes at rates $\gamma_1$ and $\gamma_2$, then by combining the two --- i.e. counting a tick from either one as a single tick --- we obtain a new Poisson process of rate $\gamma_1 + \gamma_2$. This is handy as we will show that the frequency of our reference process is a resource, the higher the better.

%% file: 4_fish.tex
\begin{figure}
    \centering
    \includegraphics[width=.9\linewidth]{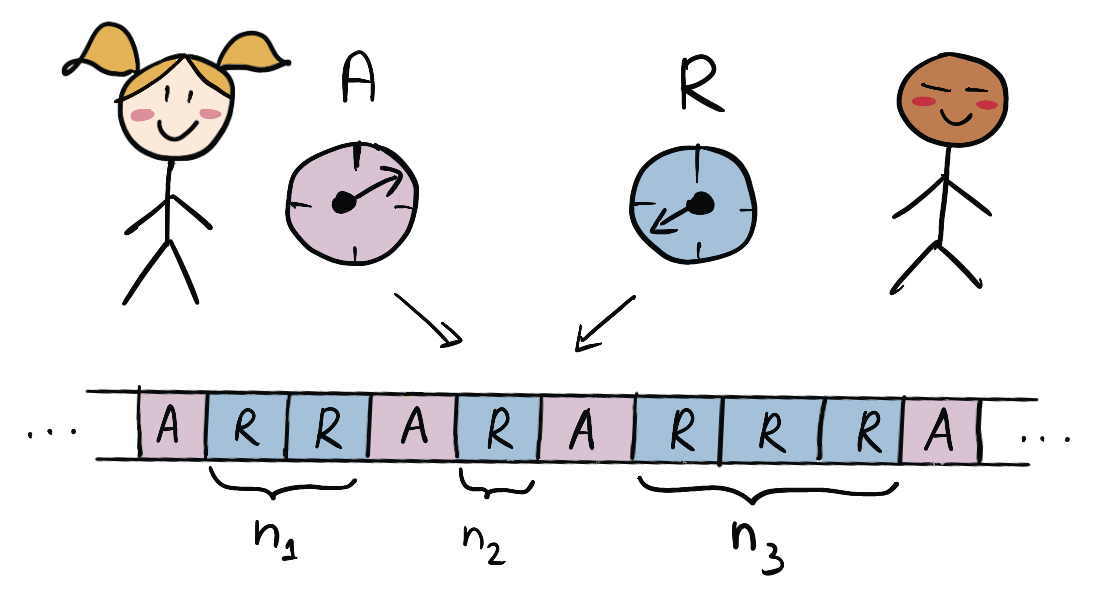}
    \caption{{\bf Measuring one clock against the other.} We characterize the clock $A$ by means of measuring it relative to a reference clock $R$, which we model as a Poisson clock. The observed statistics is then the number of ticks of $R$ between successive ticks of $A$.}
    \label{fig:clocks}
\end{figure}

Our main result concerns the following scenario: we have an imperfect clock $A$ whose ticks form an i.i.d.\ sequence so that the probability density function of multiple ticks is determined by that of a single tick:
\begin{align}
    \Pr(T_1 = t_1, &T_2 = t_2,T_3 = t_3, ...) \nonumber \\
    &= \omega(t_1) \cdot \omega(t_2 - t_1) \cdot \omega(t_3 - t_2) ...
\end{align}
$\omega(t)$ is called the waiting-time distribution, or waiting-time in short, of the clock. We characterize it by means of measuring it against a reference clock that we take to be a Poisson clock $R$ of rate $\gamma$. Our observed quantity is the number of ticks of $R$ between successive ticks of $A$; see Fig.~\ref{fig:clocks}. This itself is an i.i.d.\ variable: the probability of seeing $n$ ticks of $R$ between the $j^{th}$ and $j+1^{th}$ tick of $A$ is simply the product probability of the waiting-time being $t$ and the Poisson process probability of $n$ given parameter $\gamma t$:
\begin{align}
    \Pr(N_R = n,T_{j+1} - T_j = t) &= \Pr(N_R = n|t) \cdot \omega(t).
\end{align}

As the number of measurements increases, we can estimate the probability of observing $n$ ticks of $R$ given the clock $A$:
\begin{align}
    p_n &= \int_0^{+\infty} Pr(N_R = n | T = t) \; \omega(t)  \; dt \\
        &= \int_0^{+\infty} \frac{(\gamma t)^n e^{-\gamma t}}{n!} \omega(t) \; dt
\end{align}

We can write down the $k^{th}$ moments of the distribution $\{p_n\}_n$ (which can be interpreted as the moments of the reference $R$ relative to the clock $A$),
\begin{align}
\label{eq:relative-moments}
    m_k(\omega, \gamma) &= \sum_{n=0}^{+\infty} n^k p_n = \sum_{n=1}^{+\infty} n^k \int_0^{+\infty} \frac{(\gamma t)^n e^{-\gamma t}}{n!} \omega(t) dt.
\end{align}
To simplify this expression, we make use of the moments of the Poisson distribution $\Pr(N_R = n | T = t)$, which can be written compactly as~\footnote{For completeness, we include the derivation in Appendix~\ref{appendix:poisson-moments}.} 
\begin{align}
    M_k(t) &= \sum_{n=1}^{+\infty} n^k \Pr(N_R = n | T = t) \\&=\sum_{j=1}^k (\gamma t)^j \stirlingtwo{k}{j}, 
\end{align}
where the braces $\stirlingtwo{k}{j}$ denote the Stirling numbers of the second kind, which count the number of ways to partition a set of $k$ objects into $j$ non-empty subsets. Substituting into (\ref{eq:relative-moments}), we obtain
\begin{align}
    m_k(\omega, \gamma) &= \int_0^{+\infty} \sum_{j=1}^k \stirlingtwo{k}{j} (\gamma t)^j \omega(t) dt \\
    &= \sum_{j=1}^k \stirlingtwo{k}{j} \gamma^j M_j(\omega),\label{eq:momentsvsmoments}
\end{align}
where $M_j(\omega)$ is the $j$th moment of the waiting-time of the measured clock $A$.

Hence, having access to all relative moments $m_k(\omega, \gamma)$, we are able to reconstruct all the moments of the measured clock. The explicit expression is given by the inverse Stirling transform of the sequence $\{m_k(\omega, \gamma)\}$,
\begin{align}\label{eq:Mfromm}
    M_k(\omega) &= \frac{1}{\gamma^k} \sum_{j=1}^k (-1)^{k-j} \stirlingone{k}{j} m_j(\omega,\gamma),
\end{align}
where the square brackets $\stirlingone{k}{j}$ denote the unsigned Stirling numbers of the first kind, counting the number of permutations of $k$ elements with $j$ cycles. For example, the first and the second clock moments have the form
\begin{align}
    \gamma M_1(\omega) &= m_1(\omega, \gamma); \label{eq:firstrelative} \\
    \gamma^2 M_2(\omega) &= m_2(\omega, \gamma) - m_1(\omega, \gamma), \label{eq:secondrelative}
\end{align}
with the following moments formed in an analogous manner. The moment generating and characteristic functions can then be reconstructed as
\begin{align}
    \mathcal M(x) &= \sum_{j=0}^\infty \frac{x^j}{j!} M_j(\omega);\\
    \mathcal C(x) &= \sum_{j=0}^\infty \frac{(ix)^j}{j!} M_j(\omega).
\end{align}

The above analysis implicity assumes that the rate $\gamma$ of the Poisson process is known. However, even if $\gamma$ is unknown, one may still access the moments of the clock \textit{relative to the timescale of the reference, i.e. $\gamma^j M_j$}. One may then reconstruct the moment-generating and characteristic functions $\mathcal{M}(\gamma x)$ and $\mathcal{C}(\gamma x)$ analogously to the above. This is consistent with a fundamental fact about time-keeping: all measured time intervals are in relation to the timescale of some physical process; our current standard is the ground-state hyperfine transition frequency of the Cs-133 atom. Thus any characterisation of a clock must necessarily be expressed w.r.t.\ the timescale of another reference clock, as is the case in this work.

%% file: 5_frequency.tex
The results of the previous section show that one could recover the waiting-time distribution of an unknown clock using any Poisson process as a reference, albeit in the asymptotic limit of infinite ticks. In this section we derive error bounds on the estimates of the clock moments in the case of a finite number of ticks, proving the robustness of the results. Of specific interest is how the error depends upon the finite frequency $\gamma$ of the reference clock. An important conclusion we draw is that the error has \textit{independent} contributions from both the innate fluctuations of the clock of interest as well as the finite nature of $\gamma$.

Consider that one has a finite sample size of $N$ with elements $X_1, \dots, X_N$. Note that each sample corresponds to the interval between one tick of the unknown clock and the next, with $X_i$ the number of ticks of the reference clock in between. We can then use \eqref{eq:Mfromm} and \eqref{eq:relative-moments} to construct an estimator $\tilde{M}_k$ for the $k^{th}$ moment $M_k(\omega)$ of the waiting time of the clock,
\begin{align}
    \tilde{M}_k &= \frac{1}{\gamma^k} \sum_{j=1}^k (-1)^{k-j} \stirlingone{k}{j} \sum_{m=1}^N \frac{n_m^j}{N}
\end{align}

We analyse this estimator in Appendix \ref{appendix:sampling-higher}. It is unbiased, so that $\braket{\tilde{M}_k} = M_k(\omega)$. By calculating its variance, we can apply Chebyshev's inequality for the probability that the estimate is off by a fraction $\theta$
\begin{align}\label{eq:erroranalysis}
    Pr \left( \left| \frac{\tilde{M}_k}{M_k} - 1 \right| \geq \theta \right) &\leq \frac{1}{N\theta^2} \Big( \frac{1}{P^{(k)}} + \frac{1}{\gamma} \frac{M_{2k-1}(\omega)}{M_k^2(\omega)} \nonumber \\ & \quad \quad \quad \quad + O \left( \frac{1}{\gamma^2} \right) \Big),
\end{align}
where we have defined the ``precision of the $k^{th}$ clock moment'' as the ratio
\begin{align}
    P^{(k)} &= \frac{M_k^2(\omega)}{M_{2k}(\omega) - M_k^2(\omega)}
\end{align}
In the case of $k=1$, this corresponds to a standard precision/accuracy measure, see \cite{woods_quantum_2022,erker_autonomous_2017,yang_ultimate_2020,Silva2023,meier_fundamental_2023}, and corresponds to the inverse of the Fano Factor for stochastic processes \cite{barato_cost_2016,Silva2023}.

We can understand \eqref{eq:erroranalysis} as the confluence of multiple factors. Firstly, the product $(N\theta^2)^{-1}$ implies that the same size scales quadratically with our confidence interval $\theta$. Within the error there are --- to first order in $\gamma^{-1}$ --- two independent contributions. Firstly, the precision $P^{(k)}$: it is scale invariant, i.e. it does not change under speeding up the clock by a factor of $a$ via the transformation $\omega(t) \to a \; \omega(at)$.\footnote{If $\omega(t) \to a \; \omega(at)$, then $M_k(\omega) \to a^{-k} M_k(\omega)$} Thus $P^{(k)}$ is a measure of the fluctuations of the clock moment w.r.t. \textit{its own timescale}, and contributes to the error independently of the reference. Indeed one could repeat this analysis with a perfect reference instead of a Poissonian, in this case the error reduces to only the term involving $P^{(k)}$. The second term may be understood as a comparison between the scale of the clock moment and that of the reference, as the ratio $M_{2k-1}(\omega)/M_k^2(\omega)$ scales by $a$ under the speeding up of the clock as described above; here is where the frequency of the reference clock is important. As an illustrative example, the expression for the first moment is
\begin{align}
    \Pr \left( \left| \frac{\tilde{M}_1}{m_1} - 1 \right| \geq \theta \right) &= \frac{1}{N \theta^2} \left( \frac{1}{P^{(1)}} + \frac{\nu}{\gamma} \right), \label{eq:estimateerror}
\end{align}
where $P^{(1)}$ is the aforementioned precision and $\nu = 1/M_1(\omega)$ is the average frequency of the measured clock (there are no contributes of $O(\gamma^{-2})$ in this case). There are  thus two independent contributions to the error: the clocks own uncertainty given by $1/P^{(1)}$ and the relative frequency $\nu/\gamma$ between the measured clock and the reference Poisson process. An important consequence is that increasing $\gamma$ is useful only up to a point: once $\nu/\gamma\ll 1/P^{(1)}$ the clock's own fluctuations will dominate the error rendering a further increase in the Poisson rate superfluous, a similar statement holds for the higher moments (bounds derived in Appendix~\ref{appendix:sampling-higher}).

%% file: 6_nonreset.tex
The clock of interest has so far been taken to be \emph{irreversible} and \textit{reset}, the first implying that a tick cannot be reversed, and the second resulting in every tick being identical to the rest. While these are idealised properties, real clocks may not fulfill them perfectly. This is, for instance, the case for clocks constructed out of thermodynamic processes~\cite{barato_thermodynamic_2015,barato_cost_2016,erker_autonomous_2017}. Here, every stochastic jump present in the dynamics --- such as the one that makes the register of the clock increase by $1$ --- implies the presence of the reverse process, which cannot be set to zero unless one assumes an infinite entropy production.

For a clock whose register can move both ways, there is no notion of a time-of-arrival of a tick, but the number of ticks recorded by the register is still a well-defined quantity. Assuming that the clock is still self-timed, so that a time-independent Lindbladian operator exists to describe the behaviour of the clock,  the long-time behaviour of the moments of the number of ticks is given by polynomials in $t$:
\begin{align}
    \braket{n^j} &= \sum_{i=0}^j \alpha_i^{(j)} t^i.
\end{align}
All of the coefficients, with the exception of $\alpha_0^{(j)}$, are uniquely determined by the Lindbladian; the latter is dependent on the choice of the initial state of the clock. Thus, the long-time behaviour is entirely encapsulated by the set of coefficients $\{\alpha_i^{j}\}_{i,j}$ ($i\geq 1$), the so-called \textit{full counting statistics} of the process.

These may also be recovered using a Poisson reference clock. In this case, the tick of the target clock is not well-defined, so we rather use the ticks of the reference. One may wait for $n$ ticks of the reference clock and then measure the number of ticks $k$ of the target. By repeating this procedure, one may calculate the moments $\braket{k^j}$ (which we label by $m_j(A,\gamma,n)$) from which the coefficients $\smash{\alpha_i^{(j)}}$ are determined. The full details are found in Appendix \ref{appendix:non-reset}.

For instance, the current of ticks and the rate of variance of the current (i.e., the first two cumulants of the full counting statistics) of the target clock are estimated from
\begin{align}
    J_\infty &= \gamma \frac{m_1(A,\gamma,n)}{n}, \\
    \Sigma_\infty &= \gamma \left( m_2 (A,\gamma,n) - m_1(A,\gamma,n)^2 \left( 1 + \frac{1}{n} \right) \right),
\end{align}
from which the precision $P = J_\infty/\Sigma_\infty$ may be calculated --- this is the same as the precision introduced earlier, reducing to the earlier expression in the case of reset clocks.

%% file: 8_conclusions.tex
\paragraph{Justifying abstract information measures.} Most recent literature on ticking clocks quantifies their accuracy via the waiting-time distribution, especially the mean and variance thereof~\cite{woods_quantum_2022,erker_autonomous_2017,schwarzhans_autonomous_2021,meier_fundamental_2023,yang_ultimate_2020}. It is usually easier to deal with than operational measures such as in~\cite{rankovic_quantum_2015}. However, this is only justified if the waiting-time is recoverable from an operational scenario, and without too much difficulty. The contribution of our work is a strong positive result in that direction, placing the abstract information-theoretical characterisation of clocks performed in the literature so far on a sound operational footing.

On its own, it also represents a type of \textit{operational clock tomography}: the process of recovering the properties of a clock from measurements made w.r.t.\ to other clocks. Note that our method of reconstruction may be considered to be \textit{incoherent} and \textit{in the time domain}: incoherent, because it works via a tick register that both clocks interact with incoherently; in the time domain, because the relevant object is the sequence of ticks. In contrast, the current method of feedback in atomic clocks is coherent and in the frequency domain. There, the (laser) clock interacts coherently with the reference (Cs atoms), and the output is a measure of how well the frequencies of the target and the reference match. 

Our work thus opens up a new type of tomography for clocks. A relevant question is whether one can also perform feedback on clocks in a similar manner, i.e.\ by using the tick sequence rather than an interferometric method as is the current standard.

\paragraph{The assumption of i.i.d.} Throughout this work, we have restricted ourselves to i.i.d.\ clocks for the reference and self-timed clocks as targets. While for random processes in general this is typically a highly restrictive assumption, it is not entirely so for clocks. Firstly, there are fundamental processes that are i.i.d., such as the decay of unstable atomic and nuclear states; indeed, our current definition of the second is in terms of such a transition. Thus, the notion of an i.i.d.\ reference clock is not without merit. As for the target being self-timed, consider that this is not the case. Then one has two regimes, loosely speaking: either the drifts in the dynamics of the clock are slow w.r.t.\ its own timescale, or they are fast. 

In the former case, which includes all of the clocks that we currently construct, one may still perform tomography as described in this paper. On short timescales, this gives us a snapshot of the clock as it is; repeating it on longer timescales provides a measurement of the drift in the dynamics. This is the point of measures of frequency stability, such as the Allan variance, meant to isolate both the intrinsic ``shot noise'' and the drift.

In the latter case of fast-drifting dynamics, applying the method described here will lead to results of repeated measurements that do not converge at all, showing that the clock is, in fact, unstable.

Thus, the results described in this work are still applicable to non-i.i.d.\ targets if used carefully.

\paragraph{Operational time.} While our work has employed the notion of background Schr\"odinger time to ensconce the results, it would be interesting to translate them into a setting where time is treated more \emph{operationally}. That is, where there is no notion of a `continuous background time'; rather, the sequence of ticks from various clocks is the fundamental object from which a time reference is defined.

There are natural analogues within this picture to the background case. For instance, consider a Poisson process, which represents a clock with no memory of $t$. In the absence of background time there is no such object, however there is a relational equivalent: a Bernoulli sequence~\cite{InfoClock}. This is a collection of two or more clocks with the property that given the sequence of ticks so far, the probability that the next tick is from any one of the clocks is a fixed probability independent of the history of ticks. Thus the analogue of having a Poissonian reference $R$ would be to have a pair of clocks $Q,R$ that form a Bernoulli sequence of ticks. Either of them may then take over the role of an operational background.

Similarly, the simplest case of an unknown clock --- a reset clock --- has a relational analogue as well. We may look at the sequence of ticks formed by the clock of interest $A$ with $R$, and separately with $Q$, with probabilities $p^{(m)}_n$ and $q^{(m)}_n$ of there being $n$ ticks of $R$ (or $Q$) between the $m^{th}$ and $(m+1)^{(th)}$ tick of $A$. If these are independent of $m$, then $A$ can be said to be a reset clock w.r.t. both $Q$ and $R$.

In such a scenario we may repeat the analysis in Sec. \ref{sec:construction} to recover the moments of the clock of interest, either w.r.t. $Q$ or to $R$. However, in the absence of background time, the analysis is complicated by additional correlations. That is, unlike in the case of continuous $t$, here the moments w.r.t.\ $Q$ may not be recoverable from the moments w.r.t.\ $R$ due to the possible presence of correlations in the tick sequence of all three clocks that are not visible in any of the marginal tick sequences of a pair of them.

An interesting direction of future research is to explore this operational scenario in more detail.

\paragraph{General reference clocks.} The Poisson process suggests itself as a reference clock due to its simplicity, it is in a sense the most freely available one, in addition to being very analytically tractable as our results demonstrate. It is nonetheless interesting to consider how other references might be employed to characterise an unknown clock. For instance, in Appendix \ref{appendix:sub-poisson} we show that estimates and bounds can be constructed if the reference is \textit{sub-Poissonian} --- i.e. with a precision $P^{(1)} > 1$. Deriving sharper bounds for relevant reference clocks would be an interesting future question.

\paragraph{Relativistic generalizations.} Finally, an important future direction is to consider scenarios where the clocks are embedded in some spacetime structure: here the goal will be to reconstruct the the clock properties w.r.t.\ the spacetime background $x,t$.

%% file: appendix.tex
\section{Proof of existence of moment-generating function for finite clockworks}
\label{app:finite}

\subsection{First waiting-time of an elementary ticking clock}

To start off, we take the standard model of a ticking clock with a register: the Hilbert space is a direct product of clockwork and register spaces, the latter being a classical degree of freedom that only contains information about the number of ticks:
\begin{align}
    \tilde{\mathcal{H}} &= \mathcal{H}_C \otimes \mathcal{H}_T, \\
    \tilde{\rho}_t &= \sum_{n=0}^\infty \rho^{(n)}_t \otimes \ket{n}\!\bra{n}_T,
\end{align}
with dynamics described by a Lindbladian generated semigroup that preserves the classicality of the register, in addition to the other properties mentioned in the main text namely, that the register only moves forward and by one step at a time:
\begin{align}
    \tilde{\mathcal{L}} \left[ \tilde{\rho}_t \right] = -i \left[ H \otimes \mathds{1}_T, \tilde{\rho}_t \right] &+ \sum_k \mathcal{D}_{L_k \otimes \mathds{1}_T} \left[ \tilde{\rho}_t \right] \nonumber\\
    &+ \sum_j \mathcal{D}_{J_j \otimes \Gamma_T} \left[ \tilde{\rho}_t \right],
\end{align}
where $\mathcal{D}_A$ is the standard dissispator:
\begin{align}
    \mathcal{D}_A \left[ X \right] &= A X A^\dagger - \frac{1}{2} \left\{ A^\dagger A, X \right\},
\end{align}
and $\Gamma_T$ is the raising operator on the register,
\begin{align}
    \Gamma_T &= \sum_{n=0}^\infty \ket{n+1}\!\bra{n}_T.
\end{align}
For clarity of notation, we use $\tilde{\mathcal{L}},\tilde{\rho}$ to denote the Lindbladian and state on the joint Hilbert space of clockwork and register, with the $\mathcal{L},\rho$ denoting the same upon the clockwork space alone.

It is more useful to express the dynamics w.r.t. the clockwork states $\rho_t^{(n)}$, each corresponding to a definite number $n$ of ticks recorded by the register. To do so we define the following operators on the clockwork alone:
\begin{subequations}\label{eq:clockworkoperators}
\begin{align}
    \mathcal{L}_0 \left[ \rho \right] &= -i \left[ H, \rho \right] + \sum_k \mathcal{D}_{L_k} \left[ \rho \right], \\
    V &= \sum_j J_j^\dagger J_j, \\
    \mathcal{L}_+ \left[ \rho \right] &= \sum_j J_j \rho J_j^\dagger.
\end{align}
\end{subequations}

From these we find the equation of motion for the clockwork states to be:
\begin{align}
    d_t \rho_t^{(n)} &= \mathcal{L}_0 \left[ \rho_t^{(n)} \right] - \frac{1}{2} \left\{ V, \rho_t^{(n)} \right\} + \mathcal{L}_+ \left[ \rho_t^{(n-1)} \right],
\end{align}
where the latter term does not appear for $n=0$,
\begin{align}\label{eq:zerostate}
    d_t \rho_t^{(0)} &= \mathcal{L}_0 \left[ \rho_t^{(0)} \right] - \frac{1}{2} \left\{ V, \rho_t^{(0)} \right\}.
\end{align}
Importantly, the equation of motion for $n=0$ does not depend on any other clockwork states. This is expected as the clock is fundamentally irreversible --- once it ticks it never goes back. Thus the dynamics of each clockwork state only depends on itself and the previous ones in the sequence of states.

The initial state of the register is taken to correspond to no tick having been seen,
\begin{align}
    \tilde{\rho}_0 &= \rho_0 \otimes \ket{0}\!\bra{0}_T.
\end{align}

To find the waiting-time distribution of the first tick we may track the rate of change of the survival probability, which is the same as the probability of being in the $n=0$ state:
\begin{align}\label{eq:firstwaittime}
    \omega_t &= -d_t Q_t = -d_t \Tr \left[ \rho_t^{(0)} \right] \\
    &= \Tr \left[ \mathcal{L}_+ \left[ \rho_t^{(0)} \right] \right] \\
    &= \Tr \left[ V \rho_t^{(0)} \right].
\end{align}

\subsection{Compression of tick dynamics}

As it stands the clock Lindbladian is infinite-dimensional and hence cumbersome to work with. For the purpose of studying the first waiting-time distribution we do not require all of the dynamics, in particular the dynamics after the first tick. In view of this we construct a reduced scenario where we keep the dynamics of the first tick the same and compress the entire space of the clock after ticking into a single one-dimensional Hilbert space.

In technical terms, this corresponds to the following transformation of the Hilbert space. The original Hilbert space may be expressed as a direct sum:
\begin{align}
    \mathcal{H}_C \otimes \mathcal{H}_T = \bigoplus_n \mathcal{H}_C^{(n)},
\end{align}
where $n$ is the index of the register. We keep the $n=0$ space and compress the rest into a one-dimensional space $\mathbb{C}$, so that the new space is $\mathcal{H}_C \oplus \mathbb{C}$, a $(d+1)$-dimensional Hilbert space.

For ease of notation we will use the vector $\ket{T}$ and operator $\ket{T}\!\bra{T}$ to refer to two things depending on the expression they fall in:
\begin{enumerate}
    \item as a one-dimensional state/operator from the new space in expressions of the form $\ket{\psi} \oplus \ket{T} \in \mathcal{H}_C \oplus \mathbb{C}$,
    \item as a normalised $d+1$-dimensional state or operator thereupon with support only upon the additional space $\mathbb{C}$, in expressions of the form $\ket{T} \in \mathcal{H} \otimes \mathbb{C}$.
\end{enumerate}

Our goal is to construct dynamics upon this new space in such a way that the state at any time $t$ would be given by
\begin{align}
    \rho^{(0)}_t \oplus \left( 1 - Q_t \right) \ket{T}\!\bra{T},
\end{align}
so that the `tick' space corresponds to the part of the dynamics where the first tick has already happened.

Rather than derive the correct reduced form of the Lindbladian, we state it directly from intuition and prove that it corresponds to the correct dynamics. Starting from the three operators in \eqref{eq:clockworkoperators}, we begin by keeping the dynamics that are only upon the clockwork, simply appending the null operator on the tick space,
\begin{align}
    \mathcal{L}_0^\prime = \mathcal{L}_0 \oplus 0 \ket{T}\!\bra{T}.
\end{align}

For the tick dynamics we note that the post-tick state is no longer relevant, thus all of the required information is in the positive operator $V$ from \eqref{eq:clockworkoperators}. Given the spectral decomposition of the original $V$ as $\sum_k v_k \ket{v_k}\!\bra{v_k}$, we first modify the eigenvectors by adding a null state in the tick space
\begin{align}
    \ket{v_k^\prime} = \ket{v_k} \oplus 0 \ket{T},
\end{align}
and from these we get a set of operators $\{A_k\}_k$ that are sufficient for the tick dynamics, replacing the original set $\{J_j\}_j$:
\begin{align}
    A_k &= \ket{T}\!\bra{v_k^\prime}.
\end{align}
By constructing a dissipator from this set we find:
\begin{align}
    \sum_k \mathcal{D}_{A_k} \left[ X \right] &= \Tr \left[ V^\prime X \right] \ket{T}\!\bra{T} - \frac{1}{2} \left\{ V^\prime, X \right\},
\end{align}
where $V^\prime$ is the new form of $V$,
\begin{align}
    V^\prime &= \sum_k A_k^\dagger A_k = V \oplus 0 \ket{T}\!\bra{T}.
\end{align}

Thus the new reduced Lindbladian in full is
\begin{align}\label{eq:reducedLindbladian}
    \mathcal{L}^\prime \left[ \rho_t^\prime \right] &= \mathcal{L}_0^\prime\left[ \rho_t^\prime \right] - \frac{1}{2} \left\{ V^\prime, \rho_t^\prime \right\} + \Tr \left[ V^\prime \rho_t^\prime \right] \ket{T}\!\bra{T},
\end{align}
where each term is the direct reduction of the three operators in \eqref{eq:clockworkoperators} in the original description.

To see that the dynamics are the correct one, we note first that by construction the dynamics do not create any coherence between the tick space and the rest. By picking the initial state to be incoherent w.r.t. the direct sum,
\begin{align}
    \rho_0^\prime = \eta_0 \oplus P_0 \ket{T}\!\bra{T},
\end{align}
we can express the state at a later time $t$ in the same decomposition:
\begin{align}
    \rho_t^\prime = \eta_t \oplus P_t \ket{T}\!\bra{T}.
\end{align}

We can now straightforwardly derive the equations of motion for $\eta_t,P_t$ by writing the equation of motion \eqref{eq:reducedLindbladian} w.r.t. the decomposition:
\begin{align}
    d_t \left( \eta_t \oplus P_t \ket{T}\!\bra{T} \right) &= \left( \mathcal{L}_0 \oplus 0 \right) \left[ \eta_t \oplus P_t \ket{T}\!\bra{T} \right] - \frac{1}{2} \left\{ V \oplus 0, \eta_t \oplus P_t \ket{T}\!\bra{T} \right\} + \Tr \left[ \left( V \oplus 0 \right) \left( \eta_t \oplus P_t \right) \right] \ket{T}\!\bra{T},
\end{align}
\bigskip
which is seen to split neatly w.r.t. the direct sum into two separate equations:
\begin{subequations}  
\begin{align}
   d_t \eta_t &= \mathcal{L}_0 \left[ \eta_t \right] - \frac{1}{2} \left\{ V, \eta_t \right\}, \\
   d_t P_t &= \Tr \left[ V \eta_t \right].
\end{align}
\end{subequations}

The first of these is identical to \eqref{eq:zerostate} for the evolution of the clockwork state corresponding to no ticks \eqref{eq:zerostate}, while the second is the equation of motion for the cumulative probability of ticking --- compare to \eqref{eq:firstwaittime}. Thus if we pick $\eta_0 = \rho_0,P_0 = 0$ it will be that $\eta_t = \rho_t^{(0)}$ for all $t$, and that $P_t$ will be the probability of having ticked, $P_t = 1 - Q_t$.

\subsection{Eigenmatrix decomposition w.r.t. reduced Lindbladian}

We proceed to study the new reduced Lindbladian, using the tools from \cite{baumgartner_analysis_2008} to describe the \textit{eigenmatrices} of the Lindbladian. Since the Lindbladian is not diagonalisable in general, it will have both proper and generalised eigenvectors. We employ the following notation. Consider every instance of an eigenvalue with a proper eigenmatrix to be denoted by $\lambda_i$ and its proper eigenmatrix by $\sigma_{i,0}$, so that
\begin{align}
    \left( \mathcal{L} - \lambda_i \mathds{1} \right) \left[ \sigma_{i,0} \right] &= 0.
\end{align}
To each of these eigenvalues there may also be generalised eigenmatrices that we can take to be a Jordan chain $\{\sigma_{i,j}\}$, so that
\begin{align}
    \left( \mathcal{L} - \lambda_i \mathds{1} \right] \left[ \sigma_{i,j} \right] &= \sigma_{i,j-1},
\end{align}
where $j$ runs from $1$ to the dimension of the corresponding Jordan block of the Lindbladian when written in Jordan form. The entire set of eigenmatrices over $i,j$ form a linearly independent basis for the space of matrices, there are $(d+1)^2$ of them. Importantly, each Jordan chain is of finite size.

The evolution of each eigenmatrix under the Lindbladian operator is simple given its Jordan chain:
\begin{align}
    e^{\mathcal{L} t} \left[ \sigma_{i,j} \right] &= e^{\lambda_i t} \sum_{k=0}^j \frac{t^{j-k}}{(j-k)!} \sigma_{j,k} = e^{\lambda_i t} \theta(t)_{i,j},
\end{align}
where we have denoted by $\theta_{i,k}$ the fixed polynomial of finite degree in $t$ formed by the eigenmatrices.

By construction, the tick state is a stationary state, i.e. a proper eigenmatrix of eigenvalue $0$,
\begin{align}
    \mathcal{L}^\prime \left[ \ket{T}\!\bra{T} \right] &= 0.
\end{align}
Under the assumption that the clock eventually ticks, it is the steady state corresponding to the chosen initial state of the clock, i.e.
\begin{align}
    \lim_{t \to \infty} e^{\mathcal{L}^\prime t} \left[ \rho_0^\prime \right] &= \ket{T}\!\bra{T}.
\end{align}

Consider therefore the difference between the initial state and the stationary state in terms of the eigenmatrices of $\mathcal{L}$,
\begin{align}\label{eq:diffstate}
    \rho_0^\prime - \ket{T}\!\bra{T} &= \sum_{i,j}c_{i,j} \sigma_{i,j}.
\end{align}
Applying the evolution operator $e^{\mathcal{L}^\prime t}$ for large $t$ we see that the LHS goes to $0$.

We now argue that the eigenmatrices $\sigma_i$ that appear in the RHS of the expression must all correspond to eigenvalues with negative real part (i.e. decaying eigenmatrices). Consider that this is not the case. Then applying the evolution operator $e^{\mathcal{L}^\prime t}$ on both sides and taking $t$ large, we would be left with only the non-decaying (zero real part) eigenvalues on the RHS:
\begin{align}
    0 &= \sum_{i: \Re(\lambda_i) = 0} \sum_j c_{i,j} e^{\lambda_i t} \theta(t)_{i,j}.
\end{align}
But this is a contradiction, and thus it must be that $c_{i,j}=0$ for all $\lambda_i$ with zero real part.

Returning to \eqref{eq:diffstate}, we consider the evolution for arbitrary $t$ to get:
\begin{align}
    e^{\mathcal{L}^\prime t} \left[ \rho_0^\prime \right] - \ket{T}\!\bra{T} &= \sum_i e^{\lambda_i t} \sum_j c_{i,j} \theta(t)_{i,j}.
\end{align}
We can obtain the waiting-time distribution by applying $V$ and tracing, resulting in:
\begin{align}
    \omega_t &= \Tr \left[ V e^{\mathcal{L} t} \left[ \rho_0 \right] \right] \\
    &= \Tr \left[ V^\prime e^{\mathcal{L}^\prime t} \left[ \rho_0^\prime \right] \right] \\
    &= \Tr \left[ V^\prime \left( e^{\mathcal{L}^\prime t} \left[ \rho_0^\prime \right] - \ket{T}\!\bra{T} \right) \right] \\
    &= \Tr \left[ V^\prime  \sum_i e^{\lambda_i t} \sum_j c_{i,j} \theta(t)_{i,j} \right].
\end{align}

Since all of the $\lambda_i$ in the above expression have negative real part, the waiting-time distribution must decay exponentially, albeit multiplied by a finite polynomial in $t$. To complete the proof that the moment generating function is well-defined, we consider its definition:
\begin{align}
    M_x &= \lim_{a \to \infty} \int_0^a \omega_t e^{x t} dt \\
    &= \lim_{a \to \infty} \int_0^a \Tr \left[ V^\prime  \sum_i e^{(x+\lambda_i) t} \sum_j c_{i,j} \theta(t)_{i,j} \right] dt. \label{eq:momentfinal}
\end{align}

If we denote by $-\Delta$ the real part of the eigenvalue that is closest to zero (least negative), then for all $x \in (-\Delta,+\Delta)$ the above limit exists. The integrand is a finite sum of terms, the largest of which (in the large $t$ limit) is at worst of the form $A e^{(x-\Delta)t} t^N$, where $N$ is the largest degree of $t$ appearing in any of the polynomials $\theta_{i,j}$, and the constants $A$ are finite given the finite size of $V^\prime$ and the eigenmatrices $\sigma_{i,j}$.

\subsection{Summary}

We have demonstrated that for any combination of clock initial state and Lindbladian such that the clock definitely ticks eventually, the moment generating function $M_x$ of the waiting time distribution $\omega_t$ has a neighbourhood around $0$ where it exists, thus the moments of $\omega_t$ uniquely determine it.

We did this by proving that the state of the clock corresponding to not having ticked yet must itself decay exponentially with a rate that is at least as large as the minimal negative real part of those eigenvalues of the Lindbladian that have negative real part.

In the absence of the clockwork being finite, it need not be necessary that there is a minimal negative real part for the eigenvalues of the Lindbladian --- one may have an infinite sequence approaching $0$. However if there was such a minimal decay rate for an infinite clockwork, it may be the case that the moment generating function is well-defined, though in this case one also has to bound the now infinite sum in \eqref{eq:momentfinal} to complete the same proof.

\section{Moments of Poisson process}
\label{appendix:poisson-moments}

For completeness, here we derive of the moments of Poisson process in terms of the Stirling numbers. To do so, we need several ingredients. 

A $k$th falling power of $x$ is defined as 
\begin{align}
    x^{\underline{k}} = x\cdot (x-1) \cdot \dots \cdot (x-k+1) = \frac{x!}{(x-k)!}.
\end{align}

The Stirling numbers of the second kind $\stirlingtwo{n}{k}$ can then be used to convert usual powers to falling powers via
\begin{align}
    x^n = \sum_{k=0}^n x^{\underline{k}} \stirlingtwo{n}{k}.
\end{align}

The factorial moment of the Poisson distribution (the expectation value of $X^{\underline{k}}$) with mean $\lambda$ is equal to $\lambda^k$,
\begin{align}
    \mathbb{E}[X^{\underline{k}}]= \sum_{x=k}^\infty \frac{x!}{(x-k)!} \frac{\lambda^x e^{-\lambda}}{x!} = e^{-\lambda} \lambda^k \underbrace{\sum_{x=0}^\infty \frac{\lambda^x}{x!}}_{e^\lambda} = \lambda^k.
\end{align}
Hence, the expectation value of $X^n$ is given by 
\begin{align}
    \mathbb{E}[X^n] = \sum_{k=0}^n  \mathbb{E}[X^{\underline{k}}] \stirlingtwo{n}{k} = \sum_{k=0}^n \lambda^k \stirlingtwo{n}{k}.
\end{align}

\section{Finite sampling errors}
\label{appendix:sampling-higher}

Consider that one is estimating the $k^{th}$ moment of the unknown clock from a finite sample of size $N$, i.e. we have observed the numbers $\{n_1,...,n_N\}$ of ticks of the reference between ticks of the clock of interest. From \eqref{eq:relative-moments} in the main text, we form an estimator first for the relative moments of the reference,
\begin{align}
    \tilde{m}_k &= \sum_{m=1}^N \frac{n_m^k}{N},
\end{align}
and using \eqref{eq:Mfromm}, we obtain the estimate for the clock moment,
\begin{align}
    \tilde{M}_k &= \frac{1}{\gamma^k} \sum_{j=1}^k (-1)^{k-j} \stirlingone{k}{j} \sum_{m=1}^N \frac{n_m^j}{N}.
\end{align}

To apply Chebyeshev's inequality for this moment, we require the average and variance of the above estimator, for which we need its first two moments. The first moment is $M_k(\omega)$ itself:
\begin{align}
    \braket{ \tilde{M}_k } &= \frac{1}{\gamma^k} \sum_{j=1}^k (-1)^{k-j} \stirlingone{k}{j} \sum_{m=1}^N m_k = M_k(\omega),
\end{align}
where we have used \eqref{eq:Mfromm} again; this confirms the estimator as unbiased. It's second moment is then
\begin{align}
    \braket{\tilde{M}_k^2} &=  \frac{1}{\gamma^{2k}} \sum_{j=1}^k \sum_{r=1}^k (-1)^{2k-j-r} \stirlingone{k}{j} \stirlingone{k}{r} \sum_{m=1}^N \sum_{l=1}^N \frac{\braket{n_m^j n_l^r}}{N^2} \\
    &= \frac{1}{\gamma^{2k}} \sum_{j=1}^k \sum_{r=1}^k (-1)^{2k-j-r} \stirlingone{k}{j} \stirlingone{k}{r} \frac{1}{N^2} \left( N(N-1) m_j(\omega,\gamma) m_r(\omega,\gamma) + N m_{j+r}(\omega,\gamma) \right) \\
    &= M_k^2(\omega) \left( 1 - \frac{1}{N} \right) + \frac{1}{\gamma^{2k} N} \sum_{j=1}^k \sum_{r=1}^k (-1)^{2k-j-r} \stirlingone{k}{j} \stirlingone{k}{r} \sum_{s=1}^{j+r} \stirlingtwo{j+r}{s} \gamma^s M_s(\omega),
\end{align}
where we have used \eqref{eq:Mfromm} again to turn the sum involving $m_j(\omega,\gamma) m_r(\omega,\gamma)$ into $M_k^2(\omega)$, and \eqref{eq:momentsvsmoments} to express $m_{j+r}(\omega,\gamma)$ in terms of the clock moments.

We are primarily interested in how the errors scale w.r.t. the frequency $\gamma$ of our reference, so we keep the zeroth and first order terms, which compute to
\begin{align}
    \braket{\tilde{M}_k^2} &= M_k^2(\omega) \left( 1 - \frac{1}{N} \right) + \frac{1}{\gamma^{2k} N} \left( \gamma^{2k} M_{2k}(\omega) + \gamma^{2k-1}M_{2k-1}(\omega) + O \left( \gamma^{2k-2} \right) \right) \\
    &= M_k^2(\omega) + \frac{1}{N} \left( M_{2k}(\omega) - M_k^2(\omega) \right) + \frac{1}{N\gamma} M_{2k-1}(\omega) + \frac{1}{N} O \left( \frac{1}{\gamma^2} \right),
\end{align}
from which the variance of the estimator is
\begin{align}
    \braket{\tilde{M}_k^2} - \braket{\tilde{M}_k}^2 &= \frac{1}{N} \left( M_{2k}(\omega) - M_k^2(\omega) \right) + \frac{1}{N\gamma} M_{2k-1}(\omega) + \frac{1}{N} O \left( \frac{1}{\gamma^2} \right).
\end{align}

Chebyshev's inequality for the estimate of the $k^{th}$ clock moment thus reads:
\begin{align}
    Pr \left( \left| \tilde{M}_k - M_k \right| \geq \epsilon \right) &\leq \frac{1}{N\epsilon^2} \left( M_{2k}(\omega) - M_k^2(\omega) + \frac{1}{\gamma} M_{2k-1}(\omega) + O \left( \frac{1}{\gamma^2} \right) \right).
\end{align}
We convert it to a fractional error by expressing $\epsilon = \theta M_k$,
\begin{align}
    Pr \left( \left| \frac{\tilde{M}_k}{M_k} - 1 \right| \geq \theta \right) &\leq \frac{1}{N\theta^2 M_k^2(\omega)} \left( M_{2k}(\omega) - M_k^2(\omega) + \frac{1}{\gamma} M_{2k-1}(\omega) + O \left( \frac{1}{\gamma^2} \right) \right) \\
    &= \frac{1}{N\theta^2} \left( P^{(k)} + \frac{1}{\gamma} \frac{M_{2k-1}(\omega)}{M_k^2(\omega)} + O \left( \frac{1}{\gamma^2} \right) \right),
\end{align}
where we have denoted the ``precision'' of the $k^{th}$ moment by
\begin{align}
    P^{(k)} &= \frac{M_{2k}(\omega) - M_k^2(\omega)}{M_k^2(\omega)}.
\end{align}
The precision is a scale-invariant quantity, i.e. if one changes the waiting-time of the clock of interest from $\omega(t)$ to $a \omega(a*t)$ --- essentially speeding up the clock by a factor of $a$ --- the precision is left unaffected. On the other hand, the ratio of moments $M_{2k-1}(\omega)/M_k^2(\omega)$ scales as $a$ under such a transformation.

Note that one could perform the same analysis for estimators of $m_k(\omega,\gamma)$ rather than $M_k(\omega)$. Since the underlying distribution of the $\{X_m\}$ has moments $\{m_k(\omega, \gamma)\}$, according to Chebyshev's inequality
\begin{align}
    Pr\left(\left|\bar{X^k} - m_k(\omega,\gamma)\right|\geq \epsilon\right) \leq \frac{m_{2k}(\omega,\gamma)-m_k^2(\omega,\gamma)}{N\epsilon^2}.
\end{align}
Dividing by the rate $\gamma^k$, and substituting explicit expressions for the moments, we obtain
\begin{align}
    Pr\left(\left|\frac{\bar{X^k}}{\gamma^k} - \frac{m_k(\omega,\gamma)}{\gamma^k}\right|\geq \epsilon\right) &\leq \frac{\sum_{j=1}^{2k} \stirlingtwo{2k}{j} \gamma^j M_j(\omega) - \left(\sum_{j=1}^k \stirlingtwo{k}{j} \gamma^j M_j(\omega)\right)^2}{N\epsilon^2\gamma^{2k}} \\
    &=\frac{M_{2k}(\omega) - M_k^2(\omega)}{N\epsilon^2} + \frac{k}{\gamma}\cdot\frac{(2k-1)M_{2k-1}(\omega) - 2(k-1)M_{k-1}M_k}{N\epsilon^2} + \frac{1}{N\epsilon^2}\cdot\mathcal{O}\left(\frac{1}{\gamma^2}\right) \\
    &= \frac{M_{2k}(\omega) - M_k^2(\omega)}{N\epsilon^2} + \frac{1}{N\epsilon^2}\cdot\mathcal{O}\left(\frac{1}{\gamma}\right),
\end{align}
which is similar in form to that obtained for the estimators of $M_k(\omega)$.

\section{Tomography of non-reset clocks}
\label{appendix:non-reset}

Let us generalize the procedure for finding out the parameters of non-reset clocks. In this case, we can no longer count just the number of ticks of the reference clock $R$ between the ticks of the target clock $A$, as each tick of $A$ is not an independent event. However, we can still reconstruct the waiting time distribution of $A$ by counting the \textit{total} number of ticks of the target clock $A$ against the \textit{total} number of ticks of the reference clock $R$, which we still model as a Poisson clock.

The probability of having observed $n$ ticks of the reference clock and $k$ ticks of the target clock at any background time $t$ is given by 
\begin{align}
    P(n,k | t) = P_R(n|t)\cdot P_A(k|t).
\end{align}
Since the ticks of $R$ are i.i.d. and their waiting time is given by the exponential decay distribution $\gamma e^{-\gamma t}$, the probability of having counted $n$ ticks at time $t$ is described by the Erlang distribution
\begin{align}
    P_R(n|t) = \frac{e^{-\gamma t}\gamma^n t^{n-1}}{(n-1)!} .
\end{align}
Then the conditional probability of observing $k$ ticks of $A$ given $n$ ticks of $R$ can be written as
\begin{align}
    P(k|n) &= \int_0^{+\infty} P_R(n|t) P_A(k|t) dt \\ &= \int_0^{+\infty} \frac{e^{-\gamma t}\gamma^n t^{n-1}}{(n-1)!} P_A(k|t) dt.
\end{align}
Similarly to the consideration for reset clocks, we can write the relative moments of $A$ with respect to $R$
\begin{align}
    m_j(A,\gamma,n) &= \sum_{k=0}^\infty k^j P(k|n) \\
    &= \int_0^{+\infty} P_R(n|t) P_A(k|t) dt \\ &= \int_0^{+\infty} \frac{e^{-\gamma t}\gamma^n t^{n-1}}{(n-1)!} \sum_{k=0}^\infty k^j P_A(k|t) dt.
\end{align}

We assume that we are already operating in the regime where the target clock $A$ is in its steady state; this can be done, for example, by waiting for a sufficiently long time\footnote{this would be $t >> \lambda_{min}^{-1}$, the inverse of the eignenvalue of the Lindbladian with the smallest real part.} before starting the statistics count. In this asymptotic case, we can write $A$'s counting statistics as polynomials in $t$~\cite{schaller_technical_2014}
\begin{align}
    \langle X^j \rangle = \sum_{k=0}^\infty k^j P_A(k|t) = \sum_{i=0}^j \alpha_i^{(j)} t^i.
\end{align}
For example, the first two cumulants of the counting statistics - the asymptotic current and rate of variance of ticks - are given by
\begin{align}
    J_\infty &= \alpha_1^{(1)}, \\
    \Sigma_\infty &= \alpha_1^{(2)} - 2 \alpha_0^{(1)}\alpha_1^{(1)},
\end{align}
from which the precision of $A$ can be calculated using
\begin{align}
    R = \frac{J_\infty}{\Sigma_\infty}.
\end{align}

Substituting the polynomial into the relative moments expression, we obtain
\begin{align}
    m_j(A,\gamma,n) &= \int_0^{+\infty} \frac{e^{-\gamma t}\gamma^n t^{n-1}}{(n-1)!} \sum_{i=0}^j \alpha_i^{(j)} t^i  dt\\
    &= \sum_{i=0}^j \alpha_i^{(j)} \int_0^{+\infty} \frac{e^{-\gamma t}\gamma^n t^{i+n-1}}{(n-1)!} dt \\
    &= \sum_{i=0}^j \alpha_i^{(j)} \frac{i!}{\gamma^i} \binom{n+i-1}{i}
\end{align}

Taking the limit $n\to\infty$
\begin{align}
    \lim_{n\to\infty} \frac{m_1(A,\gamma,n)}{n} = \frac{J_\infty}{\gamma}.
\end{align}
Hence,  collecting the statistics for the relative moments, we can eventually identify the coefficients $\alpha_i^{(j)}$, which are related to individual moments of the waiting time distribution of $A$.

It is important to note that this method works not only for target clocks $A$ which have a well-defined waiting time distribution, but also for those which don't -- for example, \textit{reversible} clock, which have a probability of ticking backward~\cite{Silva2023}.

\paragraph{Error estimation.}

Using Chebyshev's inequality, if we repeat the sampling $M$ times, then for the first moment:
\begin{align}
    \Pr\left(\left|\frac{\bar X}{n} - \frac{J_\infty}{\gamma}\right|\geq \epsilon \right) &\leq \frac{\langle k^2\rangle - \langle k \rangle^2}{Mn^2\epsilon^2} \\
    &= \frac{J_\infty^2\frac{n(n+1)}{\gamma^2} + \Sigma_\infty \frac{n}{\gamma} - J_\infty^2 \frac{n^2}{\gamma^2} + \alpha_0^{(2)} - (\alpha_0^{(1)})^2 - 2\alpha_0^{(1)} J_\infty\frac{n}{\gamma}}{Mn^2 \epsilon^2} \\
    &= \frac{J_\infty^2\frac{n}{\gamma^2} + \Sigma_\infty \frac{n}{\gamma} + \alpha_0^{(2)} - (\alpha_0^{(1)})^2}{Mn^2 \epsilon^2}
\end{align}

To compare this to the error analysis in the original case (see Sec. \ref{sec:frequency}), we label by $N$ the average number of ticks of the unknown clock that pass in the whole duration of the experiment, this is given by
\begin{align}
    N &= M*n*\frac{J_\infty}{\gamma},
\end{align}
and we label by $\theta$ the fractional error in the relative frequency of the target to the reference, 
\begin{align}
    \epsilon &= \theta * \frac{J_\infty}{\gamma}.
\end{align}
After some algebra, we find that the error bound is
\begin{align}
    \Pr\left(\left|\frac{\bar X/n}{J_\infty/\gamma} - 1 \right|\geq \theta \right) &\leq \frac{1}{N \theta^2} \left( \frac{1}{P} + \frac{\nu}{\gamma} + \frac{\alpha_0^{(2)} - \left(\alpha_0^{(1)}\right)^2}{nJ/\gamma} \right).
\end{align}
This is analogous to the error in the reset case \eqref{eq:estimateerror} with the additional term due to the influence of the initial state of a non-reset clock upon the statistics. If we do have a reset clock and begin the experiment at the moment of a tick of the target clock, then $\alpha_0^{j} = 0$ for all $j$, rendering the two equations identical.

For the above expression, we note that the error is smallest for $n$ maximised. Substituting the maximum value of $n$, corresponding to having only one single sample $M=1$, we get
\begin{align}
    \Pr\left(\left|\frac{\bar X/n}{J_\infty/\gamma} - 1 \right|\geq \theta \right) &\leq \frac{1}{N \theta^2} \left( \frac{1}{P} + \frac{\nu}{\gamma} + \frac{\alpha_0^{(2)} - \left(\alpha_0^{(1)}\right)^2}{N} \right).
\end{align}

Note that multiple samples are required to estimate the higher moments, so we need $M>1$ unless we are only interested in the first moment (the frequency).

\section{Sub-Poissonian processes as a reference}
\label{appendix:sub-poisson}

The bound similar to the one discussed for the case when the reference clock is described by a Poisson distribution, applies to another class of processes called \textit{sub-Poissonian}~\cite{Ahle2021}. These are described by a distribution with the mean $\mu$ and the moment-generating function bounded by
\begin{align}
    \mathcal M(x) \leq exp\left(\mu(e^s - 1)\right) \ \forall s>0.
\end{align}
For example, if all moments of some distribution either coincide or are bounded by corresponding moments of the Poisson distribution, this distribution is sub-Poissonian.

The moments of sub-Poissonian random variables can be upper bounded by~\cite{Ahle2021}
\begin{align}
    M_k(X) \leq \mu^k \left(\frac{k/\mu}{\log\left(1+\frac{k}{\mu}\right)}\right)^k,
\end{align}
where $M_k(X) = \mathbb{E}(X^k)$ is the $k$th moment of $X$. Moreover, one can also establish a lower bound for the moments using Jensen's inequality
\begin{align}
    \left(\mathbb{E}(X)\right)^k \leq \mathbb{E}(X^k) \ \Rightarrow \ M_k(X) \geq \mu^k.
\end{align}

In our consideration, the mean is given via the average rate of the process $\gamma$, that is, $\mu=\gamma t$, for the relative moments of the measured clock with delay function $\omega(t)$ relative to the sub-Poissonian clock the bounds read
\begin{align}
    \int_0^{+\infty} (\gamma t)^k \omega(t) dt \leq m_k(\omega,\text{sub-Poiss.})  \leq \int_0^{+\infty} (\gamma t)^k \left(\frac{k/\gamma t}{\log\left(1+\frac{k}{\gamma t}\right)}\right)^k \omega(t) dt.
\end{align}
The lower bound is proportional to the $k$th moment of the measured clock; to simplify the upper bound, we use that for all $y>0$, $\log(1+y) \geq \frac{y}{1+y}$,
\begin{align}
    \frac{k}{\log\left(1+\frac{k}{\gamma t}\right)} \leq k + \gamma t,
\end{align}
and the final bound on the relative moments reduced to
\begin{align}
    \gamma^k M_k(\omega) \leq m_k(\omega,\text{sub-Poiss.}) \leq \int_0^{+\infty} (k + \gamma t)^k \omega(t) dt = \sum_{j=0}^k \binom{k}{j} \gamma^j k^{k-j} M_j(\omega).
\end{align}

This allows us to derive the bound on the variance of $k$th relative moment for the sample of size $N$ with the sample average $\bar X$,
\begin{align}
    P\big(\left|\frac{\bar {X^k}}{\gamma^k} - \frac{m_k(\omega,\text{sub-Poiss.})}{\gamma^k}\right| \geq \epsilon\big) &\leq \frac{m_{2k}(\omega,\text{sub-Poiss.}) - m_k^2(\omega,\text{sub-Poiss.})}{N\gamma^{2k}\epsilon^2} \\
    &\leq \frac{1}{N\gamma^{2k}\epsilon^2} \left(\sum_{j=0}^{2k} \binom{2k}{j} \gamma^j (2k)^{2k-j} M_j(\omega) -\gamma^{2k} M_k^2(\omega) \right) \\
    &= \frac{M_{2k}(\omega) - M_k^2(\omega)}{N\epsilon^2} + \frac 1 N \cdot \mathcal{O}\left(\frac{1}{\gamma}\right),
\end{align}
where, similarly to the Poisson-distributed process, in the limit of $\gamma\to +\infty$ the relative measurement of the moments is as good as the direct one (as given by Chebyshev's inequality).